\documentclass[12pt,titlepage]{article}
\usepackage{amsmath}
\usepackage{amsfonts}
\usepackage{amssymb}
\usepackage{amsthm,amsmath}
\usepackage{amssymb,latexsym}
\usepackage{amscd,hyperref}
\usepackage[noend]{algpseudocode}
\usepackage{algorithm}
\usepackage{breqn}
\usepackage{diagbox}
\usepackage{enumitem}
\usepackage{cite}
\usepackage{geometry}
\makeatletter
\renewcommand{\fnum@algorithm}{\fname@algorithm}
\makeatother
\usepackage{color}
\usepackage{graphicx}

\title{Optimal Responses to an Infectious Disease}
\author{Evangelos F. Magirou\thanks{The author thanks Professor Spyros Skouras of the Department of International and European Economic Studies, Athens University of Economics and Business for constructive discussions and pointing out reference \cite{NBERw26981}.  Professor Nicos Christodoulakis of the same Department made extensive and incisive comments  throughout the course of this work.  
		Professor Stavros Toumpis, Department of Informatics, also of the Athens University of Economics and Business pointed out an error in the original formulation of the probability condition stated in \eqref{ProbAssum}. 
		Of course errors and omissions are the sole responsibility of the author.}\\\small{Professor Emeritus, Department of Informatics}\\\small{Athens University of Economics and Business}\\\small{Patission 76, Athens, Greece}\\efm@aueb.gr}
\date{First Version: May 26, 2020\\This Version: June 3, 2020}
\begin{document}

\maketitle

\begin{abstract}
We analyze an optimal control version of a simple SIRS epidemiology model.  The policy maker can adopt policies to diminish the contact rate between infected and susceptible individuals, at a specific economic cost. The arrival of a vaccine will immediately end the epidemic.  Total or partial immunity is modeled, while the contact rate can exhibit a (user specified) seasonality.  The problem is solved in a spreadsheet environment.  A reasonable parameter selection leads to optimal policies which are similar to those followed by different countries. A mild response relying on eventually reaching a high immunity level is optimal if ample health facilities are available.  On the other hand, limited health care facilities lead to strict lock downs while moderate ones allow a flattening of the curve approach.
\end{abstract}

\section{Introduction}
The balance between measures to reduce the spread of a virus and the desire to retain social and economic activity to a reasonable level is quite delicate, as evidenced by the bitter debate among policy makers, political parties and nations defending their own attitudes while vilifying those of others.  The tradeoffs can be assessed by using a control methodology, as in \cite{NBERw26981}, where a lock-down intervention is incorporated in a SIR model, including the probabilistic occurrence of a vaccine, \cite{Cowles2229} where properties of the optimal policy are proven,  \cite{Charpentier2020} which employs this methodology to assess the combination of several modes of intervention, while \cite{RePEc:eie:wpaper:2004} incorporates an SIR model without extensive use of optimal control methodology.

We are interested to study policy related questions  using optimization methods and in particular to clarify the conditions under which a relaxed policy is optimal versus a more restrictive one.  We are in the midst of a crisis and several questions have been  only partially answered by authorities in the Northern Hemisphere concerning:
\begin{enumerate}
	\item  To what extent will the lock-down be relaxed during the summer months? 
	\item Will the epidemic flare up again in the winter?
	\item  Given that significant immunity has not been achieved in most countries, will the intervention be more (less) intensive in the Fall?
\end{enumerate}
Our analysis shows that the policies followed by the authorities can be correct for the appropriate combination of epidemiological and economic parameters.   It is of interest to see whether the policies derived in our simple model appear in more complex deterministic or stochastic models.  Needless to say an accurate parameter estimation as well as state identification is of paramount importance. Policy making under imperfect information is beyond the scope of this work.
\section{Model Formulation}\label{Model}
We consider the standard SIRS epidemiological model  (without vital dynamics) of W. O. Kermack and A. G. McKendrick \cite{Kermack},  \cite{HethcoteSIAM} endowed with the potential for controlling the contact parameter. Thus let $S(t),I(t),R(t)$ be the number of individuals in a population of size $N(t)$ that are  \textit{Susceptible, Infectives and Removed} at time $t$.  We consider a short horizon, assume a constant population $N=N(t)$ and work with the corresponding fractions $s(t)=S(t)/N$ and $v(t)=I(t)/N$, the removed fraction being determined by  $s,v$ as $1-s(t)-v(t)$.  For convenience we denoted the fraction of infectives by \textbf{$v$} instead of \textbf{$i$}.  

The dynamics of an epidemic  assuming homogeneous mixing of the population are determined by: 

a. The rate of newly infected originating in the susceptibles is $\lambda(t)s(t)v(t)$ 

b. The rate of removal from the class of infectives is $\gamma v(t)$ and 

c. The rate of immunity loss of those removed is  $\delta (1-s(t)-v(t))$ who then revert to the susceptible class

In the above equations $\lambda$ is the contact rate, which is the average number of adequate contacts per infective and unit time \cite{HethcoteSIAM},  $\gamma$ is the removal (recovery plus death) rate. The case of a nonzero reinfection rate corresponding to immunity loss at rate $\delta$ can be easily included in the model (and causes no numerical difficulties) so we are dealing with a controlled form of the SIRS model. The standard epidemic dynamics are 
\begin{equation}\label{SIR}
\begin{array}{l}
\frac{ds}{dt}=-\lambda(t)s(t)v(t)+\delta(1-s(t)-v(t))\\
\frac{dv}{dt}=\lambda(t)s(t)v(t)-\gamma v(t).
\end{array}
\end{equation}
 For the infectives to decrease it is necessary that $s(t)<\frac{\gamma}{\lambda}$, the quantity $\frac{\gamma}{\lambda}$ being referred to as the \textit{basic reproduction number} $R_o$ \cite{HethcoteSIAM}.  Equivalently one reports \cite{ReprodNumbers} the \textit{Effective Reproduction Number} $R_e=\frac{\lambda s(t)}{\gamma}$  which is desired to be less than one to control the epidemic. Equilibria of \eqref{SIR}   have been analyzed extensively \cite{HethcoteSIAM} assuming no loss of immunity.  More detailed models in the same reference \cite{HethcoteSIAM} include latent infectives (Exposed) who accelerate the contagion rate (SEIRS). We assume that this effect can be reflected by a proper choice of the contact rate $\lambda$ without drastically altering the optimal intervention policies.  We allow for the contact rate to depend on time, modeling thus seasonal variations in the spread of viruses.
 
 We assume that the mitigation, suppression and any other policies mentioned say in \cite{Imperial} can be represented by a single, scalar control variable $u(t)$ in $[0,1]$, as in \cite{NBERw26981}. This is an oversimplification: one could study modes of intervention $k=1,..,n$ and consider a vector control $[u_1,..,u_n] $ of intensities of the corresponding instrument as in \cite{Charpentier2020}.  However all proposed interventions consist of reducing the contact rate $\lambda$ so we expect to get some insights from the scalar case.  We thus model the effect of a level $u$ intervention  by the expression
\begin{equation}
\lambda(t)=\lambda_o(t)(1-u(t)).
\end{equation}
The contact rate before the intervention is $\lambda_o(t)$ and most likely has a seasonality.  The effectiveness of the intervention is assumed linear multiplicative, while \cite{NBERw26981} assumes a quadratic effect $\lambda(t)=\lambda_o(t)(1-\theta u(t))^2$ with $0\le\theta \le1$ the effectiveness of the intervention.  This formulation assumes that the intervention is of the lockdown type and hence there will be a reduction in both the number of infectives that are free to move and the number of contacts each one has.  We prefer to use a linear multiplicative effect which simplifies calculation while there little loss of generality since the two formulations are functionally related.  Note that by considering a $\theta = \theta(t)$ that is increasing, even exceeding unity (as long as the contact rate is nonnegative) one could model increasing intervention efficiency.

At every time interval $[t,t+\Delta t]$ we assume a cost proportional to the fraction of infectives.  The cost will consist of the reduced output of those infected that exhibit symptoms, the cost of medical service required (perhaps in addition a discomfort cost) and finally a cost for fatalities.  In \cite{NBERw26981} the dimensions of this cost are in economic terms, and we will follow their estimates in our parameter selection.  Thus we will have a cost element $m(t)v(t)\Delta t$, the time dependence in the cost coefficient $m(t)$ possibly reflecting changes in treatment costs and effectiveness. We also include a cost element that reflects the capacity of a health system as follows: let $v_o(t)$ be the level of infectives that causes the heath system to reach its capacity; it can be time dependent reflecting capacity additions.  Then we add  a smooth penalty cost element of the form $v(t)\cdot ae^{M(v(t)-v_o)}\Delta t$. The parameters $a,M$ are selected so that the cost is small if the infectives do not stress the system but rises steeply if they do. In contrast, \cite{NBERw26981} models this capacity problem by assuming a cost quadratic in the fraction of infectives $v$; a non smooth penalty term of the form $M\cdot max[0,v-v_o]$ with $M$ large is introduced for the same purpose in \cite{Charpentier2020}.  

The control intensity $u$ is assumed to incur a cost which is convex since simple mitigation policies have sub linear costs, but suppression measures have costs that increase super linearly.  Thus we consider a control cost of the form $Au(t)^n \Delta t$, usually quadratic $(n=2)$. We will follow again \cite{NBERw26981} in selecting parameters, although it does not include a quadratic control cost.  We could also incorporate a cost term reflecting weariness from lengthy interventions, an important concern in policy making; this term could be a function of the total effort up to that time, say of  $z_t=\int_{0}^{t} u(s)ds$.  However the higher dimension of such a model would make its numerical solution too involved for a preliminary analysis. 

The prospective of finding an effective cure including a vaccine has been a major concern in the planning of interventions.  We assume that a complete cure will be available at a time $T_V$ at which the crisis ends and no further cost accrues, not even a terminal one.  It is straightforward  to introduce such a cost that depends on the final level of susceptibles and infectives but this would again unduly complicate the numerical solution. 

The random variable $T_V$ has a density $g_{t_o}$ and the corresponding probability measure $P_{t_o}$ which are assessed at the initial time $t=t_o$, and which change as new information becomes available, so that at $t_1>t_o$ the perspectives for a vaccine are now expressed by the density $g_{t_1}$ and the measure $P_{t_1}$.  We will find useful to deal with the tails distribution $G_{t_o}(t) \equiv P_{t_o}(T_V\ge t)=\int_{t}^{\infty}g_{t_o}(s)ds$ for $t\ge t_o$.  

Assuming now that costs are additive in time and are to be discounted at a rate $\rho$ the total cost is given by the random variable:
\begin{equation}\label{totcost}
\int_{t_o}^{T_V}e^{-\rho \tau}\left[ v(\tau)\left\lbrace m(\tau) +a\exp(M[v(\tau)-v_{o}])\right\rbrace+Au^n \right] d\tau.
\end{equation}
With $g,G$ as before, the expected value of the random variable in \eqref{totcost} is 
\begin{equation}\label{CostRV}
\int_{t_o}^{\infty}g_{t_o}(T_V)\int_{t_o}^{T_V}e^{-\rho \tau}\left[ v(\tau)\left\lbrace m(\tau) +a\exp(M[v(\tau)-v_{o}])\right\rbrace+Au^n \right] d\tau dT_V.
\end{equation}
We will assume that the cure will  be found by a finite time $T_{max}$ and so $g_{t_o}(t)$ vanishes for $t$ greater than $T_{max}$. Then the domain of integration in \eqref{CostRV} is finite and so changing the order of integration we write the expected cost as
\begin{equation}\label{Objective}
\int_{t_o}^{T_{max}}e^{-\rho \tau}G_{t_o}(\tau)\left[ v(\tau)\left\lbrace m(\tau) +a\exp(M[v(\tau)-v_{o}])\right\rbrace+Au^n \right] d\tau.
\end{equation}
It is important to observe that if the vaccine has not been found by some time $t_1>t_o$ the minimization problem is essentially different from that in \eqref{Objective}  since the integrand now  includes the factor $G_{t_1}$  instead of $G_{t_o}$ and thus a control $u^*(t_1)$ that has been established optimal at $t_o$ does not retain this property. This undesirable feature will not occur  if the statistics of the vaccine arrival time are invariant in the following sense: At any time $t_1$ before the end of the epidemic, \textit{we assume that the vaccine arrival  statistics used at $t_1$ are those estimated at the initial time $t_o$ conditioned on the event that a vaccine has not been developed by $t_1$.}  Thus in terms of the tails distribution we have for $ t\ge t_1\ge t_o$:
\begin{equation}\label{ProbAssum}
G_{t_1}(t)\equiv P_{t_1}(T_V\ge t)=P_{t_o}(T_V\ge t\mid T_V \ge t_1)=\frac{P_{t_o}(T_V\ge t)}{P_{t_o}(T_V\ge t_1)}\equiv \frac{G_{t_o}(t)}{G_{t_o}(t_1)}
\end{equation}  
This assumption is expressed in the second equality in \eqref{ProbAssum},

The expected cost to go at time $t_1>t_o$ is 
\begin{equation}\label{Objectivet1}
\int_{t_1}^{T_{max}}e^{-\rho \tau}G_{t_1}(\tau)\left[ v(\tau)\left\lbrace m(\tau) +a\exp(M[v(\tau)-v_{o}])\right\rbrace+Au^n \right] d\tau
\end{equation}
which given assumption \eqref{ProbAssum} becomes
\begin{equation}\label{Objectivet2}
\frac{1}{G_{t_o}(t_1)}\int_{t_1}^{T_{max}}e^{-\rho \tau}G_{t_o}(\tau)\left[ v(\tau)\left\lbrace m(\tau) +a\exp(M[v(\tau)-v_{o}])\right\rbrace+Au^n \right] d\tau.
\end{equation}
However this expression is just a constant dividing the part of the cost integral \eqref{Objective} starting at $t_1$ and ending at $T_{max}$.  Since this integral must also be a minimum by the principle of optimality, the control determined at $t_o$ remains optimal until the end of the horizon.

It is often assumed in the literature that at every time instant the vaccine arrival time has a negative exponential distribution (e.g. \cite{NBERw26981}, \cite{Charpentier2020}).  This is consistent with assumption \eqref{ProbAssum} since if $g_{t_o}(t)=\psi \exp(\psi(t-t_o))\quad t\ge t_o$ then the tails distribution $G_{t_o}(t)=\exp(-\psi (t-t_o))$  satisfies $G_{t_1}(t)=G_{t_0}(t)/G_{t_o}(t_1)$.  Therefore the policies determined assuming a negative exponentially distributed vaccine arrival time do not require any adjustment. 

As customary, we try to determine the policy $u(t)$ that minimizes the expected value of the cost \eqref{Objective} subject to the dynamics in \eqref{SIR}, a standard optimal control problem.  Using assumption \eqref{ProbAssum} we can drop the $t_o$ index of $G$ in \eqref{Objective}.  In addition to minimizing expected value one could consider utility criteria that take into account various statistics of the cost random variable, possibly a portfolio type problem minimizing variance given a upper bound on the expected cost. These problems are more complicated so we work with expected value minimization. Since our formulation is time dependent we use the optimal control formalism \cite{Bryson};  a solution through the Hamilton Jacobi Bellman equation used in \cite{NBERw26981} would have an additional time dimension and will be difficult to solve.  We therefore consider the Hamiltonian
\begin{align}\label{Hamiltonian}
\begin{split}
H=e^{-\rho \tau}G(\tau)\left[ v(\tau)\left\lbrace m(\tau) +a\exp(M[v(\tau)-v_{o}])\right\rbrace+Au^n \right]+\\+\phi_s \left[  \lambda_o(t)(u(t)-1)s(t)v(t)+\delta(1-s(t)-v(t))\right] +\\+\phi_v \left[  \lambda_o(t)(1-u(t))s(t)v(t)s(t)-\gamma v(t)\right]. 
\end{split}
\end{align}
The costate variables are $\phi_s,\phi_v$ for states $s,v$ respectively.  The optimal policy is determined by choosing the value of $u$ in $[0,1]$ minimizing the Hamiltonian \eqref{Hamiltonian}.  The Hamiltonian is convex in the control so we solve $\frac{\partial H}{\partial u}=0$ for $u^*$, 
\begin{equation}\label{OptContr1}
u^*(t)=\left[ \frac{e^{\rho t}(\phi_v-\phi_s)\lambda_o(t)s(t)v(t)}{nAG(t)}\right] ^{1/(n-1)}
\end{equation}
and obtain the optimal control by truncation: 
\begin{equation}\label{OptContr}
u_{opt}(t)=\left\lbrace
\begin{array}{lr}
1&u^*\ge 1 \\
u^* &0\le u^*\le 1\\
0& u^*\le 0
\end{array}\right.
\end{equation}

\noindent The optimal intervention expression above shows from a policy perspective why it is important to have a good estimates of the product of susceptibles and infectives. In \cite{Cowles2229} properties of the optimal policy are derived for a constant underlying contact rate $\lambda_o(t)=\lambda_o$ and it would be of interest to extend their analysis to the case where we have a seasonal variation in the rate.

The costate equations are
\begin{equation}\label{Costateequ}
\begin{array}{l}
\frac{d\phi_s}{dt}=-\frac{\partial H}{\partial s}= (\phi_s-\phi_v)\lambda_o(t)(1-u(t))v(t)+\delta \phi_s\\ \frac{d\phi_v}{dt}=-\frac{\partial H}{\partial v}=(\phi_s-\phi_v)\lambda_o(t)(1-u(t))s(t)+\delta \phi_s+\gamma  \phi_v-\\\;\;-\left[m(t)+a(1+v(t)M)\exp(M v(\tau)-v_{o}(t))\right] e^{-\rho t}G(t).
\end{array}
\end{equation}
The optimal control is determined by solving a two point boundary value problem \textbf{(TPBVP)} consisting of the equations \eqref{SIR} and \eqref{Costateequ} with control as in \eqref{OptContr}.  The boundary conditions are (a) at the initial time $t_o$ we are given the values of the state variables $s(t_o)=s_o,\; v(t_o)=v_o$ (b) at the final time $T_{max}$ it is the costate variables that must vanish, $\phi_s(T_{max})=\phi_v(T_{max})=0$. 

Such \textbf{TPBVP's} are notoriously difficult to solve numerically (a comprehensive description is in Ch. 7 of \cite{Bryson}) since the costate variables increase backward in time while the state ones decrease, and indeed we encountered numerical instabilities.  Our computations used several methods described in the reference, usually starting with a variation of the shooting method \cite{Conte} that consists of selecting arbitrary starting values for $\phi_s(t_o),\phi_v(t_o)$ along with the given $s(t_o),v(t_o)$ and solving the initial value problem to obtain the corresponding final costate values $\phi_s(T_{max}),\;\phi_v(T_{max})$.  The initial value problem of the four differential equations \eqref{SIR} and \eqref{Costateequ} with conditions at $t_o$ was solved by a second order predictor corrector method \cite{Conte} in a spreadsheet environment.  We then adjust the initial costate values until $\phi_s(T_{max}),\;\phi_v(T_{max})$ vanish. This is numerically difficult because small changes in the initial costate values cause huge ones in the final ones \cite{Bryson}.   In particular, we used the built in spreadsheet optimizer to adjust the initial costate variables so that the final ones vanish. When this did not give satisfactory results (final values were far from zero) we adjusted the initial values to minimize cost, which often decreased the policy cost but led to large final costate values.  In such cases we improved on the solution by the first order gradient search for optimal control problems (Section 7.4 in  \cite{Bryson}) which is also used in \cite{Charpentier2020}.  Usually the improvement was slight and the derivative $\partial H/\partial u$ decreased slowly, so we mostly report on the results of the shooting method. A detailed description of the numerical methods used (and spreadsheets) is available upon email request. 

\section{Parameter Selection}\label{SecParEst}
We select nominal values for the parameters informally, since these values only serve as reference points around which we vary them to form scenarios. In turn these  will show the types of optimal intervention we expect to see implemented.  We will select these nominal parameter values to be consistent with those stated in the literature and then modify them in various scenarios.

To select epidemiological parameters we use mainly \cite{NBERw26981} and \cite{Imperial}.  The contact rate we use is potentially periodic of the form $$\lambda_o(t)=\lambda_o (1+k_{seas}\sin (2\pi t)).$$  The time unit is one year, and the origin $t=0$ is the Autumn Equinox. Thus in late September the contact rate is $\lambda_o$ but in the beginning of the winter it takes the value $\lambda_o(1+k_{seas}).$  The nominal value for the seasonality coefficient is set at $0.8$, but we also tested smaller nonzero values.  The base contact rate $\lambda_o$ is selected as in \cite{Imperial} to reflect a $20\%$ daily increase or (about) 70 $year^{-1}$.  The removal rate $\gamma$ is taken as $1/18\;days^{-1}$ or about $20\;year^{-1}$.  The mortality rate is taken as in \cite{Imperial} $\mu =1\%$, so the number of infection caused fatalities in an interval $\Delta t$ is $\mu \gamma Nv(t)\Delta t$.  The fatality rate is age dependent and thus models to be used for detailed policy making should include age related variables; here we restrict ourselves to a homogeneous population.  The fatality rate has been challenged in \cite{Ioannidis} since the infectives' population was estimated to be quite higher than the official  one. We will study this variation in some scenarios.

To assess the economic impact of the infection we make the following assumptions: it is stated in \cite{Imperial} that about $60\%$ of the infected show some symptoms, the rest being asymptomatic or unreported (we will not differentiate between the two).  Of those exhibiting symptoms (which appear on the average 5 days after infection)  a fraction of $0.04$ need hospitalization, and $0.30$ of the above require intensive care. We thus have four types of infectives, those who show No Symptoms, those with Mild Symptoms, those with Severe Symptoms and those with Extra Severe Symptoms.  The types of health care are also of four types, No Care, Home Care, Hospital Care and ICU Care.  In Table \ref{Paramcomp} we (loosely) assign fractions of infection time to care type for each infective type.  As an example, those who will show Severe Symptoms will spent 28\% of their infection time asymptomatic, 28\% in home care and 44\% in a hospital.  Thus a randomly selected infective will be in a specific state of health care with a probability calculated in the above table by the expression relating conditional to total probability, i.e. $$P(Care Type)=\sum_{InfectiveType}P(Care Type\mid InfectivexType) P(InfectiveType).$$
Thus the probability of being in Home Care is $0.577\cdot 0.50+.016\cdot0.28+0.007\cdot0.11 = 0.294$.  This calculation is shown in the last row of Table \ref{Paramcomp}. 

\begin{table}	
	\centering
	\caption{Disease parameters}  
	\label{Paramcomp}   
	\begin{small}
		\begin{tabular}{|c|c|c|c|c||c|}
			\hline
			Infective type	&	No symptoms& Home&	Hospitali-&Intensive&Fraction \\	
           && Care&zation&Care&\\	
			\hline
		No symptoms&	1.00&	0.00	&	0.00	&	0.00 & 0.400	\\
			Mild Symptoms	&	0.50	&	0.50	&	0.00 & 0.00 & 0.577	\\
			Severe Symptoms	&	0.28	&	0.28	&	0.44	& 0.00 & 0.016 \\
			Extra Severe	&	0.17	&	0.11	&	0.39 & 0.33 & 0.007	\\
			\hline
			
			\hline
			Probability	&	0.694	&	0.294	&	0.010 & 0.002 & 1.000	\\
			\hline
		\end{tabular}
	\end{small} 
\end{table}

 We compute an economic cost of the infection consisting of production loss plus health care costs as follows.  In an interval of length $\Delta t$ those symptomatic will abstain from production and will cause additional costs to the health system depending on the symptoms' severity.  We assume a homogeneous population of $N$ individuals producing W units of a single good yearly,  thus lost production is $(1-0.694)Nv(t)\frac{W}{N} \Delta t=0.306v(t)W\Delta t)$.  We assume that the daily cost of home treatment is half, hospital treatment is 5 times and intensive care 20 daily production units. Then average treatment cost is the fractions of care types derived earlier multiplied by the corresponding costs, namely $$(0.294\cdot 0.5+0.010\cdot5+0.002\cdot20)v(t)W\Delta t.$$ Adding the cost of foregone production we obtain the expression  $$(0.306+0.294\cdot 0.5+0.010\cdot5+0.002\cdot20)v(t)W\Delta t=0.543v(t)W\Delta t.$$

 In the analysis of \cite{NBERw26981} the cost of fatalities is expressed in terms of production lost, which is $\frac{W}{rN}$ if we assume individuals of infinite life, where $r,N$ are the discount factor and population respectively.  The same analysis considers  adding a lump sum $\chi$ as a noneconomic loss of life cost, which however is taken as zero.  Now, individuals do not live forever and furthermore fatalities in this particular infection are mainly among older individuals of lower life expectation so $\frac{W}{rN}$ is an overestimate.  We thus use the same figures as in \cite{NBERw26981}, i.e. a cost $20W/N$ per fatality but which implies an additional noneconomic fatality cost.  Loss of life in $\Delta t$ is $\mu \gamma v(t)\Delta t$ so the fatality cost is $0.01\cdot 20 \cdot 20 v(t)\Delta t = 4 v(t)\Delta t$.  This term is an order of magnitude greater than the production foregone and health care costs, and so exact estimation of the latter is of minor importance. Naturally, a revision of the fatality rate will greatly influence the total infection cost.  In most scenarios the coefficient of the infected fraction is taken as 5, but we also examine a value of 2.5, reflecting a lower fatality rate as implied in \cite{Ioannidis}.
 
 The model incorporates a steep penalty term when the capacity of the health system is exceeded consisting of adding the term $ae^{M(v(t)-v_o)}$ to the infectives' coefficient $m$, $v_o$ being the infected fraction that drives the health system to its capacity. We select $a$ to be unity and $M$ to be large, about 200, and we will verify the appropriateness of this choice.  The critical level of  infectives is selected taking into account that a fraction of 0.002 of those infected require intensive care (Table \ref{Paramcomp}, last row) and the proportion of ICU units are $10^{-4}$ in a typical western country (the case of Italy as recounted in \cite{Imperial}).  Thus setting $2\cdot10^{-3}v_o=10^{-4}$ we obtain $v_o=5\%$. For zero infected the additional cost coefficient is $\exp(-200\cdot 0.05)=4.5\cdot10^{-5} <<5$; for $v-v_o=2\%$ the added coefficient is $\exp(200\cdot 0.02)=54.5>>5$; for $v=v_o$ the addition to the nominal coefficient of $5$ is 1, i.e. 20\%, which reflects some stress on the health system. We will also consider additional capacity to the health system by increasing $v_o$ as $v_o(t)=v_o2^t$, doubling the capacity every year (one could consider a construction cost, but this would unduly complicate a preliminary model).
 
 The  cost  $Au^n,\;n>1$ is a  convex function of the intervention level, which is reasonable since first order results can be achieved without any production loss through warnings, information dissemination etc. Suppression policies do have an economic cost, indeed a total stop of the contagion $u=1$ can only be achieved with strict isolation and hence an almost total loss in production.  This in turn would translate in loss of life, so we feel justified in keeping $W$ the yearly production  as the accounting unit without expressing it in terms of fatalities imputed to economic hardship.  This means that the value of $A$ should roughly correspond to the value of total loss of production and in utility terms should be greater than one.  An alternative way to select $A$ with a quadratic cost is to estimate a level of intervention whose cost is numerically equal to its effectiveness that is $Au_o^2=u_o$ and $A=u_o^{-1}$.  Then for $u \le u_o$  the cost is below the effectiveness of the intervention and conversely for higher values.  For instance, if we have cheap intervention only for contagion reductions smaller than $10\%$ (a situation where intervention is difficult) this implies a high value of $A=10$.  We will examine scenarios with various values for $A$, even lower than unity.  
 
 The arrival of a successful cure and/or a vaccine is considered a random variable with known density. We mainly used a uniform distribution in $[1,2]$ reflecting the widespread belief that a vaccine will be available in 18 months. The arrival  probability is considered negative exponential in \cite{NBERw26981} and \cite{Charpentier2020} but the prevailing view is that it's impossible to find such a cure in at least a year, which is incompatible with the exponential distribution.  

 \section{Types of Optimal Intervention}\label{OptInterv}
We have calculated the optimal policies for several  parameter sets we refer to as scenarios.  These policies are presented mostly  diagrammatically.  We show in the Figures for each policy the trajectory of the susceptibles and the control, both being in the unit interval. \textit{We  also show the trajectory of the infectives fraction scaled by the capacity level $v_o(t)$ so most of the time $v(t)/v_o(t)$ is in $[0,1]$.} 

For reasonable variation in parameters we generate optimal policies that  correspond to those that have been implemented by the countries inflicted by the virus, the moral being that any of the policies actually followed is reasonable for a specific parameter selection.  The particular parameters we have used are shown in Table \ref{Paramscenarios}, where we have indicated by boldface entries differing from those in the previous row.  In each row  we show for the corresponding scenario  health system capacity, cost of intervention, cost of infection per time unit, and reinfection rate. In the second part of the table we show some computational results, namely total cost of the scenario, the initial and final values of the costate variables. We omitted the final values of the state variables since these are shown in the corresponding diagrams.  In all scenarios the initial values are $v(0)=0.1 \%$ of the population being infected, $s(0)=98\%$  susceptible, the rest being immune.   The penalty exponential in \eqref{totcost} has a coefficient of $M=200$ and unit multiplier $a$. Other parameters that are common in all cases are: Contact rate $\lambda_o=70\;year^{-1}$, Recovery rate $\gamma=20\; year^{-1}$, Discount rate $\rho=3\%$, Seasonality $k_{seas}=80\%$. A perfect cure will occur in a random time uniformly distributed between one and two years, this being the most widely expressed experts' estimate.  Variation in these parameters will be shown in the appropriate scenarios. We ran several variations of the above scenarios, altering the discount rate and the seasonality parameter; the results did not substantially differ from the main scenarios and will not be presented.
\begin{table}	
	\centering
	\caption{Scenario parameters and results}  
	\label{Paramscenarios}   
	\begin{small}
		\begin{tabular}{|c|c|c|c|c||c|c|c|c|c|c|}
			\hline
		Scen-&Bound on&Interven-& Infec-&Rein-&Sce-&Initial&Initial&Final &Final\\	
		ario &Infected&tion Cost& tives  &fection&nario&$\phi_s$&$\phi_v$&$\phi_s$&$\phi_v$ \\	
		          No.&$v_o$ & A &Cost  $m$ & Rate $\delta$ &Cost& & & & \\	\hline
1	&	1.00	&	0.5	&	5	&	0.0	&0.243	&	0.254	&0.213&	0.008	&	1.081	\\	\hline
2	&	1.00	&	0.5	&	5	&\textbf{0.3}&0.252	&	0.243	&0.416&	0.014	&	0.000	\\ \hline
3	&	\textbf{0.05}&\textbf{2.0}&	5	&\textbf{0.0}&	1.165&	0.598	&25.63&	0317&-1.302	\\\hline
4 	&	0.05	&\textbf{20.0}&	5	&0.0&4.531&	  5.159 &120.14&1.992&	1384.5	\\ \hline
5 	&	0.10	&\textbf{2.00}&\textbf{2.5}&0.0&0.411&	0.478	&10.71&	4.604&-1.4$\cdot10^8$\\	\hline
\end{tabular}
\end{small} 
\end{table}

\begin{figure}[ht]
	\centering
	\includegraphics[width=12cm, height=7.5cm]{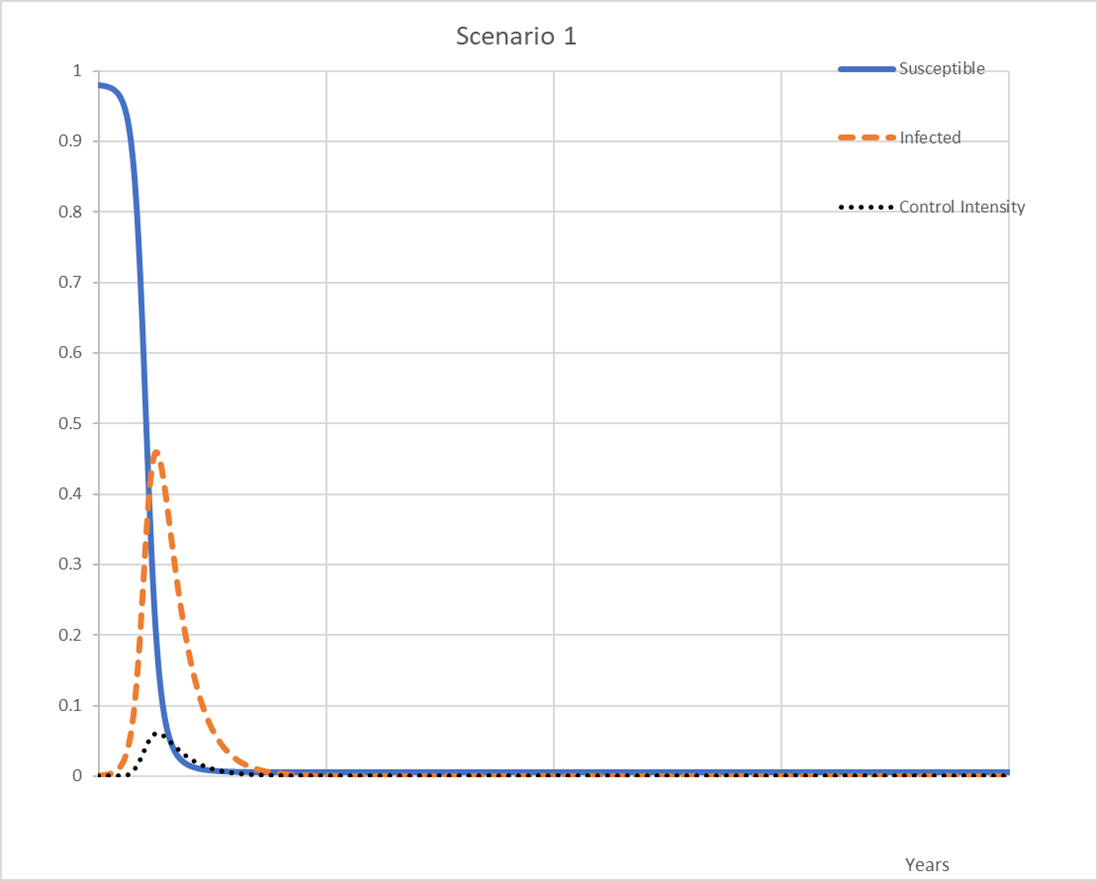}
	\caption{\small{\textbf{High capacity system, Small intervention cost, no reinfection} } }
	\label{Case_1}
\end{figure}

We first examine a situation of a high capacity health system for which there is no bound on the infected fraction, and zero reinfection.  The intervention cost is low with a value of $A=0.5$, i.e. vanishing of the contact rate can be achieved with a production loss of 50\%.  This is Scenario 1 in the Table, and the corresponding Figure \ref{Case_1}. Note that the final costate variable in the Table is somewhat large (about unity) so we applied the gradient method to improve on the policy: this did not significantly alter it, giving a cost improvement of just $10^{-6}$ confirming the results in the Table.  In this policy it is optimal to impose light measures and essentially let the infection take its course.  The maximum infected will be in three months and the total population will be in the removed state by then. An undesirable effect is that we have a high infected fraction (almost half of the population) at a single instance; this might cause a social disruption that has not been properly accounted in the cost structure. The total cost is calculated at $0.243$ which was expected: The total time spent by the infectives in this state is $1/20$ years, at cost $5$, so the total cost for the susceptibles (98\% of the population) is $5\cdot0.98/20=0.245$ which is slightly larger than the optimal cost of $0.243$. Note also that the infectives' fraction will always be below 50\%. 

\begin{figure}[ht]
	\centering
	\includegraphics[width=12cm, height=7.5cm]{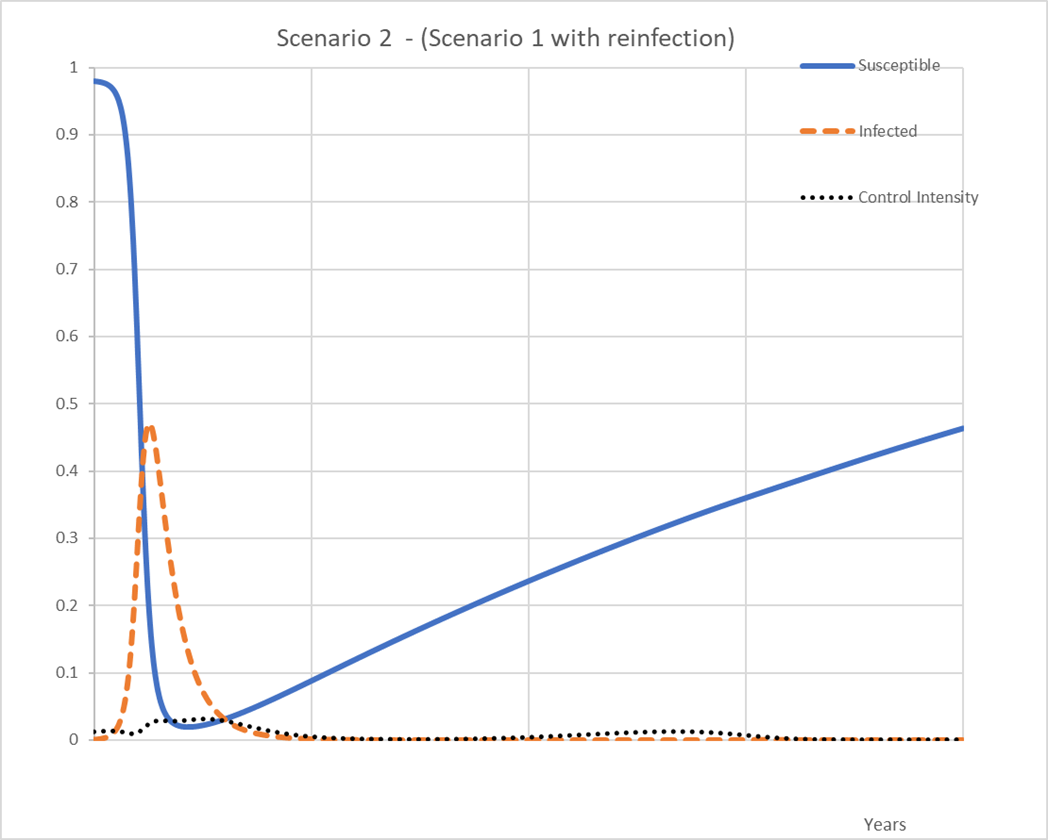}
	\caption{ \small{\textbf{High capacity system, Small intervention cost, Reinfection}} }
	\label{Case_2}
\end{figure}

We assume now that there is a loss of immunity with 30\% of these removed becoming susceptible in one year. Thus in Scenario 2, Figure \ref{Case_2} we have an optimal policy consisting of slightly more intense measures and almost the same cost $0.252$, but a strongly rising population of susceptibles until the horizon end at year 2.  Again, at $t=0.25$ a high fraction of  the population is infected, with a serious danger of social disruption. For $t\ge0.25$ susceptibles are mostly below $\gamma/\lambda=20/70 \approxeq 0.29$ which is required for the infectives' fraction to decrease. 

\begin{figure}[ht]
	\centering
	\includegraphics[width=12cm, height=8cm]{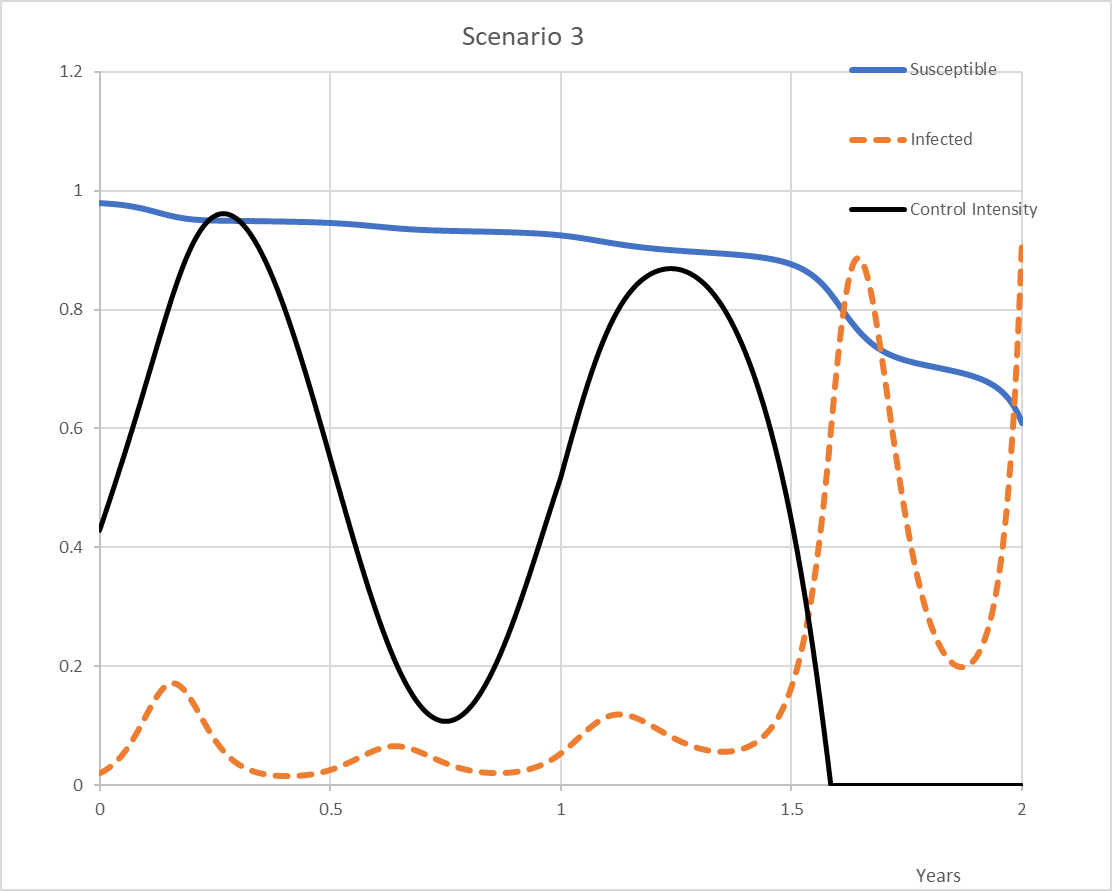}
	\caption{\small{\textbf{Low capacity system, Medium intervention cost, No Reinfection}} }
	\label{Case_3}
\end{figure}

We next consider health systems that will be stressed by a high level of infectives. In line with the parameter selection of Section \ref{SecParEst} we consider a maximum of infectives $v_o$ to be either 5 or 10 percent.  We include the possibility of expanding the health system without extra cost, so we consider either a constant $v_o$ or of  the form $v_o(t)=v_o 2^t$, a characteristic that will be shown to have  little effect on policy since the increase arrives too late.  In Scenario 3 we consider a barely adequate health system with infectives capacity $v_o=5\%$, a reasonable intervention cost of $A=2$ and no capacity expansion. The optimal policy consists of implementing strong measures to keep infection fraction to acceptable levels, see Figure \ref{Case_3} (again we exhibit $v/v_o$ instead of $v$). This is a \textbf{strict suppression policy} where infectives remain low and susceptibles high throughout the period, and thus requires reimposing measures in the  next autumn.  Intervention permanently ceases after $t\approxeq1.5$  in the expectation of the arrival of the vaccine.  As a result infectives start increasing and reach the capacity level before starting to drop due to the seasonal reduction in the contact rate.  They start to rise again at $t\approxeq1.75$ but fail to exceed capacity by the end of the horizon.  The cost of this Scenario is almost five times that of  scenarios 1 and 2 (1.17 versus only 0.25), as was to be expected given the higher control cost and the inferior health system.  Computationally, although the final costate variable are reasonably close to zero, applying the gradient method gives a similar policy with 3\% lower cost, so we kept the result in the table.  A similar policy is optimal even in the case of reinfections (not shown), which is to be expected as susceptibles stay high anyway.  A capacity doubling also yields a similar policy, the infected being now a lower fraction of capacity.  Increasing the arrival horizon of the vaccine by assuming a uniform distribution in [1,3]  also calls for intensive measures  keeping the infected population low and stopping close to  the horizon's end.  We wrongly conjectured that a herd of immunity policy would have been better given the longer horizon; this might be true for a longer one.  

\begin{figure}[ht]
	\centering
	\includegraphics[width=12cm, height=7.5 cm]{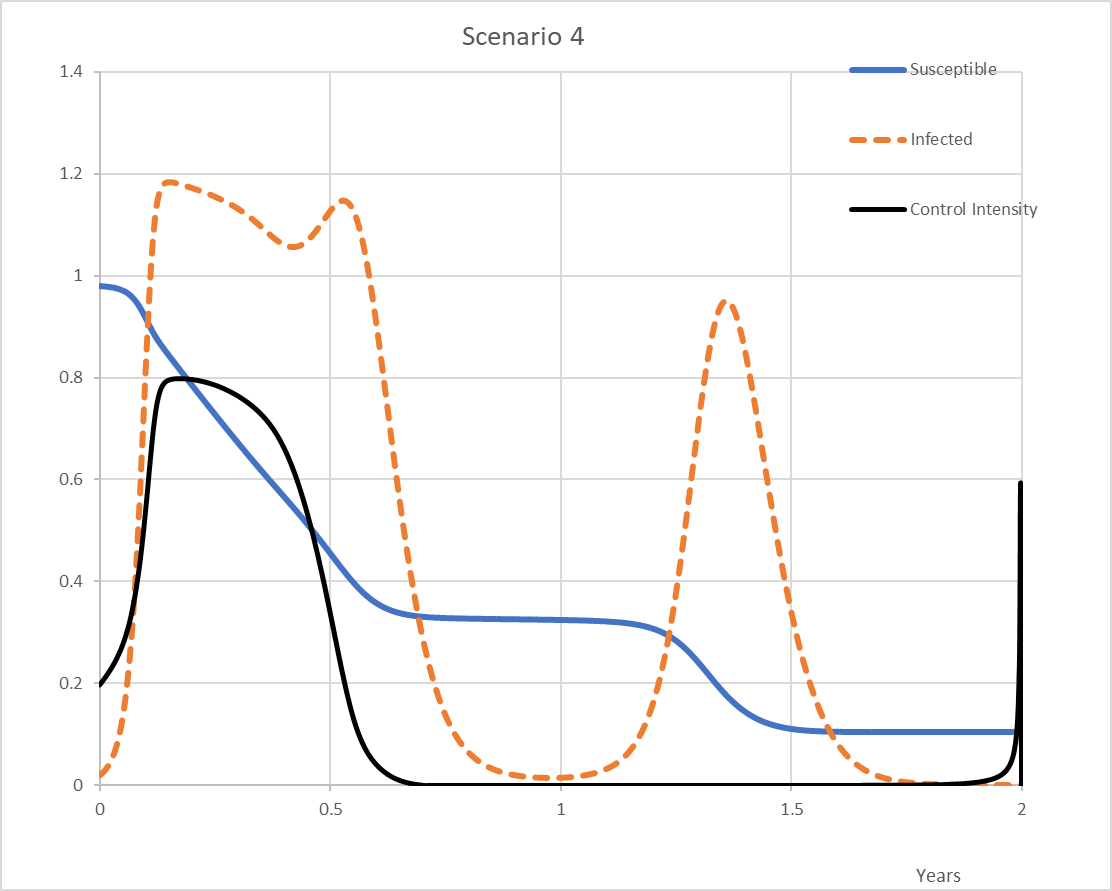}
	\caption{\small{ \textbf{Low capacity system, High intervention cost, No Reinfection - Flattening the Curve Policy}} }
	\label{Case_4}
\end{figure}

\begin{figure}[ht]
	\centering
	\includegraphics[width=12cm, height=7.5 cm]{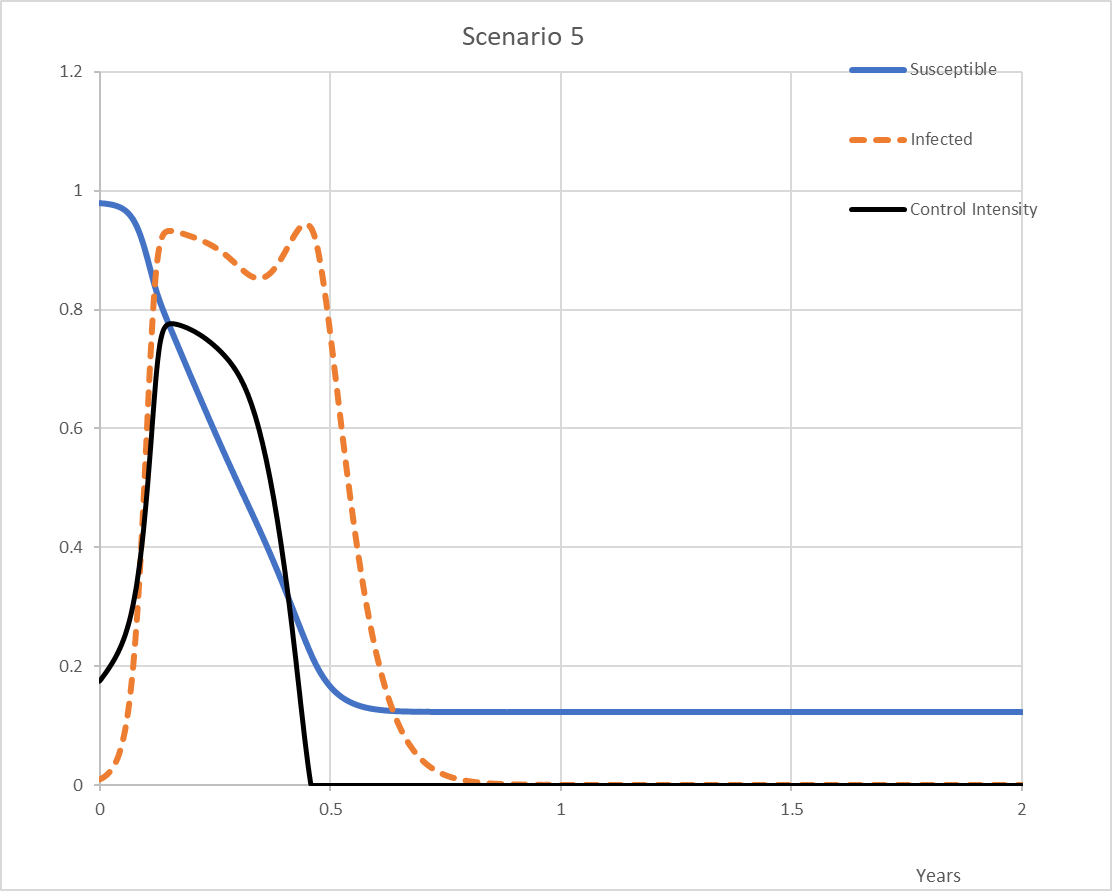}
	\caption{ \small{\textbf{Medium capacity system, Medium intervention cost, No Reinfection - Flattening the Curve Policy}}} 
	\label{Case_5}
\end{figure}

Increasing  the cost of  intervention significantly $(A=20)$ alters the character of the optimal policy  - this is Scenario 4 in the Table and Figure \ref{Case_4}.  The measures are strict in the beginning but are soon interrupted at $t\approxeq 0.5$.  The infected exceed the upper bound by 20\%, so the solution is not strictly acceptable although the results are instructive.  The measures do not repeat in the following year so the infectives rise, but given the smaller susceptible fraction of $s\approxeq0.4$  they do not exceed the bound $v_o$ while the susceptibles decrease further. This is consistent with what is popularly referred as \textbf{flattening the curve strategy}, allowing sufficient immunity to be attained at a rate that does not exceed the capacity of the health system. The optimal policy for a longer horizon has a similar structure.  From the numerical point of view, given the large values of the final costate values, it is possible to further decrease the cost by the gradient method.  This improvement is again small (about 4\%) with a similar policy, so we kept the original results. We do not have an intuitive justification for the control bump towards the horizon's end, in contrast with Scenario 3 where measures cease as the arrival of the vaccine approaches.  
A policy that also is of the flattening the curve - type occurs for a medium cost of intervention $(A=2)$, moderately prepared health system (the fraction of infected $v_o$ can be up to 10\%) while the  cost of infection is also moderate reflecting optimistic assessments of the virulence \textit{(m=2.5)}.  This is Scenario 5 in Table \ref{Paramscenarios}, and the optimal response is shown in Figure \ref{Case_5}. Again the final values of the costates are high, so we applied the gradient method, which again gave insignificant improvements (less than 1\%, a cost of 0.409 instead of 0.411 shown in the Table), so we kept the calculations in the Table.  The optimal policy again consists of a strong intervention during the winter (from $t=0$, the autumn equinox to $t=0.5$ the spring equinox) peaking at $u=.8$ and stopping not to be repeated.  The infected population slightly exceeds  capacity but then decreases, shows a second peak right after the end of the measures and then vanishes under the combined effect of the seasonal drop in the contact rate and the reduction in susceptibles to about $15\%$ (which is lower than the level guaranteeing the eradication of the infection $20/(1.8*70)=15.9\%$).  This policy is optimal even for smaller intervention costs but at a very low cost $(A=1)$ it is optimal to implement a strict intervention as in Scenario 3.  We also examined the effect of reinfection, and it proved be minor: if in Scenario 5 we assume that 30\% of those recovered lose their immunity in one year the optimal policy changes only slightly. 

As of the time of writing, measures are being substantially relaxed throughout Europe.  This can be considered as a tacit acceptance of either a strong seasonality in the contact rate, or a reassessment of the cost of intervention.  Furthermore, estimates of the acquired immunity in most countries are well below 30\% of the population.  This seems to indicate that a type of Scenario 3 policy is being followed, and hence (as has been stated by spokespersons of the public health authorities) we should expect some sort of repetition of the interventions towards the end of the year.  Given the experience acquired by public health authorities we expect more effective interventions - a feature that can also be incorporated in a model that includes learning by doing.

\section{Model extensions, conclusions}
The model could be enhanced in several directions:
\begin{itemize}
	\item Introduce more population classes, for instance age groups, symptomatic and asymptomatic infectives as done in \cite{Charpentier2020}
	\item Consider spatially distributed models, leading to diffusion partial differential equations which have a similar optimal control treatment
	\item The states $s,v$ have to be estimated in any stochastic SIR type model.  A Kalman filtering approach on a linearization of the SIR equations with imperfect observation and noise could be easily implemented, complementing the ubiquitous data analysis methods that have been proposed.
	\item Investigate the characteristics of an optimal policy for variable underlying contact rates, using techniques as in \cite{Cowles2229}
\end{itemize}

\bibliography{BibCovid}

\begin{thebibliography}{10}

\bibitem{NBERw26981}
Fernando~E Alvarez, David Argente, and Francesco Lippi.
\newblock A simple planning problem for \uppercase{covid-}19 lockdown.
\newblock Working Paper 26981, National Bureau of Economic Research, April
  2020.

\bibitem{ReprodNumbers}
Jeffrey~K Aronson, Jon Brassey, and Kamal~R. Mahtani.
\newblock When will it be over?: An introduction to viral reproduction numbers,
  \uppercase{R}o and \uppercase{R}e.
\newblock
  https://www.cebm.net/covid-19/when-will-it-be-over-an-introduction-to-viral-reproduction-numbers-r0-and-re/,
  19 April 2020.
\newblock Centre for Evidence-Based Medicine, Nuffield Department of Primary
  Care Health Sciences, University of Oxford.

\bibitem{Ioannidis}
Eran Bendavid, Bianca Mulaney, Neeraj Sood, Soleil Shah, Emilia Ling, Rebecca
  Bromley-Dulfano, Cara Lai, Zoe Weissberg, Rodrigo Saavedra-Walker, James
  Tedrow, Dona Tversky, Andrew Bogan, Thomas Kupiec, Daniel Eichner, Ribhav
  Gupta, John Ioannidis, and Jay Bhattacharya.
\newblock Covid-19 antibody seroprevalence in \uppercase{S}anta
  \uppercase{c}lara \uppercase{c}ounty, \uppercase{c}alifornia.
\newblock {\em medRxiv, Cold Spring Harbor Laboratory Press}, 2020.

\bibitem{Bryson}
Arthur Bryson and Yu-Chi Ho.
\newblock {\em Applied Optimal Control}.
\newblock Ginn and Company, 1969.

\bibitem{Charpentier2020}
Arthur Charpentier, Romuald Elie, Mathieu Lauriere, and Viet~Chi Tran.
\newblock \uppercase{Covid}-19 pandemic control: \uppercase{b}alancing
  detection policy and lockdown intervention under \uppercase{Icu}
  sustainability.
\newblock arXiv:2005.06526v3 [q-bio.PE], May 13 2020.

\bibitem{Conte}
Samuel~Daniel Conte and Carl W.~De Boor.
\newblock {\em Elementary Numerical Analysis: An Algorithmic Approach}.
\newblock McGraw-Hill Higher Education, 3rd edition, 1980.

\bibitem{Imperial}
Neil~M Ferguson, Daniel Laydon, and Gemma Nedjati-Gilani~et al.
\newblock Impact of non-pharmaceutical interventions (\uppercase{NPI}s) to
  reduce \uppercase{covid-19} mortality and healthcare demand.
\newblock https://doi.org/10.25561/77482, March 20, 2020.
\newblock Imperial College London.

\bibitem{Kermack}
W.O. Kermack and A.~G. McKendrick.
\newblock A contribution to the mathematical theory of epidemics.
\newblock {\em Proc. Roy. Soc.}, A115:700--721, 1927.

\bibitem{Cowles2229}
Thomas Kruse and Philipp Strack.
\newblock Optimal control of an epidemic through social distancing.
\newblock Discussion Paper 2229, Cowles Foundation for Research in Economics,
  Yale University.

\bibitem{RePEc:eie:wpaper:2004}
Facundo Piguillem and Liyan Shi.
\newblock Optimal \uppercase{COVID-19} quarantine and testing policies.
\newblock EIEF Working Papers Series 2004, Einaudi Institute for Economics and
  Finance (EIEF), 2020.

\bibitem{HethcoteSIAM}
Hethcote.~H. W.
\newblock The mathematics of infectious diseases.
\newblock {\em SIAM Review}, 42:599--653, 2000.

\end{thebibliography}
\bibliographystyle{Plain}
\end{document}